# All-optical switching of a dye-doped liquid crystal plasmonic metasurface

**Bernhard Atorf, Holger Mühlenbernd, Thomas Zentgraf, and Heinz Kitzerow***

*Center for Optoelectronics and Photonics Paderborn (CeOPP), Paderborn University, Warburger Str. 100, 33098 Paderborn, Germany*
*\*heinz.kitzerow@upb.de*

**Abstract:** A switchable metasurface composed of plasmonic split ring resonators and a dye-doped liquid crystal is developed. The transmission of the metasurface in the infrared spectral range can be changed by illuminating the dye-doped liquid crystal with light in the visible spectral range. The effect is particularly efficient in the case of hybrid alignment of the liquid crystal, i. e. alignment of the director perpendicular to the surface on one substrate and parallel alignment on the counter substrate. This all-optical switching effect can be attributed to the behavior described in earlier works as colossal optical nonlinearity or surface-induced nonlinear optical effect.

## 1. Introduction

The interaction of waves with materials that contain functional structures with sub-wavelength size (metamaterials) can facilitate many unusual phenomena. Metamaterials [1-38] can be described by macroscopic effective material parameters. Their unusual properties can be attributed to extreme values [1] or to a tailored spatial distribution of these effective material parameters [2, 3]. Since the prediction of negative refractive indices [4] and their experimental realization [5], many fundamental studies have been performed and possible applications have been demonstrated, which include perfect lenses [6], hyperlenses [7], optical cloaks [8], etc. The fabrication of bulk metamaterials is challenging [9]. However, metasurfaces – two-dimensional arrays of sub-wavelength structures – are easier to fabricate and in many cases sufficient to provide the function of a frequency-selective filter, a polarizer, a lens or a similar optical element by merely a very thin structured layer [10 – 38]. Since plasmonic resonances of metals applied in many earlier metamaterials or metasurfaces are known to be associated with large absorption and thus optical losses, more recent studies concentrated on the development of dielectric or semiconducting Mie resonators [11-19] or metal-dielectric hybrid structures [20-26]. For practical applications, it is important to fabricate tunable structures that can be controlled by external stimuli [15-38], thereby facilitating the development of active components like – for example – optical switches or modulators, adjustable filters or lenses, transducers and sensors. Recently developed tuning methods are based on micro- or nanoelectromechanical systems [15, 16], charge carrier density changes in semiconductors [17-18, 21], amorphous-crystalline phase changes in dielectrics [22, 23] or metal–insulator transitions in $VO_2$ [24-26]. A more traditional way of achieving addressable optical properties is based on the use of a liquid crystal (LC), which serves as a dielectric host with an effective permittivity that is sensitive to temperature [27] and external magnetic [28] or electric fields [27, 29-38]. As the name indicates, LCs combine properties of liquids and crystals. They are ordered liquids, which can be birefringent although they are fluid. Due to their unique properties, LCs became ubiquitous in electro-optical applications. The most commonly used LCs consist of rod-like organic molecules and are referred to as calamitic LCs. LCs can possess several mesophases with different degrees of order. Very frequently, the nematic phase appears. In the nematic phase, the LC is uniaxially birefringent due to the preferred parallel orientation of neighboring molecules. The average orientation of the LC molecules can be described by a pseudovector – the director **n** –, the direction of which may depend on the position in a given sample. The director **n** is influenced by the anchoring conditions of the confining substrates, the elastic properties of the LC and external stimuli, like applied magnetic or electric fields [39]. A uniaxially birefringent LC acts as a homogeneous retarder when the director is uniformly aligned. If so, the LC influences the polarization state of transmitted radiation depending on the azimuthal angle between the plane of polarization and the optical axis. However, if the director **n** is twisted between two confining substrates with planar alignment, a twisted nematic (TN) structure appears [40-42]. The latter behaves similar to a stack of gradually twisted mica sheets [43], which cause a rotation of the plane of polarization of linearly polarized light. This waveguiding effect is, under certain conditions, almost independent of the wavelength. For photonic data processing, it is essential to control optical properties by light through nonlinear optical effects. However, most studies on tunable metasurfaces utilizing LCs have focused on controlling the director by quasi-static electric fields, so far.

In order to explore the opportunity of optical addressing, we describe here the influence of the optical Kerr effect of a nematic LC on the transmission of a metasurface composed of plasmonic split ring resonators (SSRs). We have chosen this plasmonic platform because SRRs [Fig. 1(a)] are easy to fabricate, flat and relatively robust and theoretically well understood.

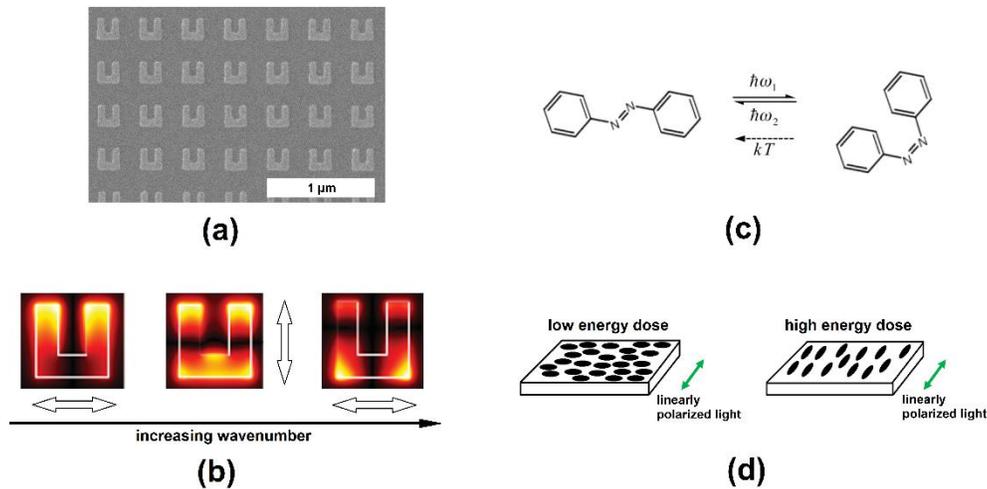

Fig. 1. (a) Scanning electron microscopy image of the split ring resonators studied. (b) Shape of plasmonic split ring resonators and spatial distribution of the electric field amplitudes for modes excited by infrared radiation with different planes of linear polarization. Arrows indicate the electric field direction of the exciting radiation. (c) Cis-trans isomerization of the dye methyl red. (d) Phenomenon of the light-induced realignment of a liquid crystal that appears owing to the colossal optical nonlinearity [34] and is known as the surface-induced nonlinear effect (SINE) [33].

SSRs were among the first sub-wavelength structures, which were designed for achieving negative values of the permeability and of the refractive index owing to electromagnetic resonance. Since the experimental demonstration of this function in the microwave range [44], they have been systematically scaled down to operate also at infrared (IR) frequencies [45]. SRRs have been frequently characterized and are very well known. A study in the IR range revealed distinct resonant modes, which can be described by a quantum number corresponding to the number of nodes of the spatial distribution of the induced electric polarization [46]. These resonant modes can be selectively excited by polarized light. If the gap of the SRR points upwards [Fig. 1 (b)], odd-numbered (even-numbered) modes are excited by linearly polarized radiation with a horizontal (vertical) plane of polarization. The nearfields at the resonances shown in Fig. 1 (b) were calculated by a finite difference time domain (FDTD) method. Within this method, electrical currents inside the gold structure were simulated assuming a Drude polarizability. In the desired frequency, the parameters of the Drude model were adjusted to match the dielectric functions of gold as published by Johnson and Christy [47]. The calculations were done with periodic boundary conditions and a mesh size of 1 nm.

The nonlinear optical properties of an LC are used to control the state of polarization of incident IR radiation, thereby facilitating changes in the effective transmittance of the metasurface. Nematic LCs are known to show a giant optical Kerr effect [48, 49], i. e. a very strong dependence of the effective refractive index

$$n = n_0 + n_2 \cdot I \qquad (1)$$

on the intensity I of the incident light. In usual nematic LCs, the nonlinear coefficient $n_2$ (Table 1) can exceed the respective value of a typical isotropic fluid considerably.

Table 1. Orders of magnitude of the nonlinear index of refraction $n_2$ (cf. equation 1) observed in thermotropic nematic liquid crystals (data from Ref. [49]).

| Material | Order of magnitude of $n_2$ [cm²/W] |
|---|---|
| Pure nematic LC | $10^{-4} – 10^{-3}$ |
| Usual dye-doped nematic LC | $10^{-3} – 10^{-1}$ |
| Nematic LC doped with methyl-red | 1 - 2000 |

Dye-doped nematic LCs show even larger nonlinear coefficients $n_2$ (Janossy effect [50]). For the azo dye methyl red (MR) [Fig. 1 (c)], a particularly large value of $n_2$ was found (colossal nonlinear effect [51]). The latter effect has been extensively studied under different experimental conditions (sample thickness, surface anchoring, light intensities) [52-54]. Thorough studies revealed a surface-induced nonlinear effect (SINE) in addition to the bulk effect [54, 55]. The former was attributed to adsorption and desorption of dye molecules at the surface. It was found that the light-induced alignment of the rod-like LC molecules depends essentially on the energy dose of illumination. Small (large) energy doses of irradiation with linearly polarized light induce an alignment of the LC director perpendicular (parallel) to the plane of polarization [Fig. 1 (d)].

## 2. Experiment

The SRRs made of gold were fabricated by standard electron beam lithography and subsequent thermal deposition of 30 nm gold with the lift-off of the mask material PMMA. Both the lateral dimension and the spacing of the SRRs are 200 nm [Fig. 1(a)]. For this size of the SRRs, the first two resonances appear at wavenumbers between 4000 and 9000 cm$^{-1}$, as described earlier [27, 29]. This metasurface was combined with an MR-doped (5 wt. %) liquid crystal (5CB) and optically addressed. Before studying the response of the SRRs, baseline investigations were performed to characterize the performance of the MR-doped LC with respect to the SINE described in previous works. Subsequently, the LC-combined SRRs were studied.

The near-infrared (NIR) intensities and spectra were measured in a Fourier transform IR spectrometer, type Equinox 55 from Bruker (Germany) using a tungsten filament lamp with a coupled Hyperion 1000 microscope from Bruker (Germany). Additionally, a visible laser beam ($\lambda$ = 520 nm, P = 30 mW, laser type LDCU8/A336 from Power Technology) was coupled into the beam path in order to optically address the samples [Fig. 2]. For beam expansion, a 10x achromatic beam expander (Thorlabs GBE10-A) was used. To merge the IR-beam and the laser beam, a cold mirror (Thorlabs FM203) was placed at a tilt angle of 45° in the beam path. The cold mirror was mounted in a kinematic mount (Thorlabs KM200T) and placed on a displacement rail (smalltec ARC-15) in order to facilitate its optional use. To decrease the attenuation of the laser beam, most of the gold mirrors were replaced with silver or aluminum mirrors. Both the NIR radiation and the visible laser beam were linearly polarized by means of an ultra-broadband wire grid polarizer ($\lambda$ = 250 nm – 4 µm, Thorlabs WP25M-UB). For opto-optical baseline investigations of the MR-doped liquid crystal, a NIR linear polarizer (Edmund optics #48-888) was used as an analyzer.

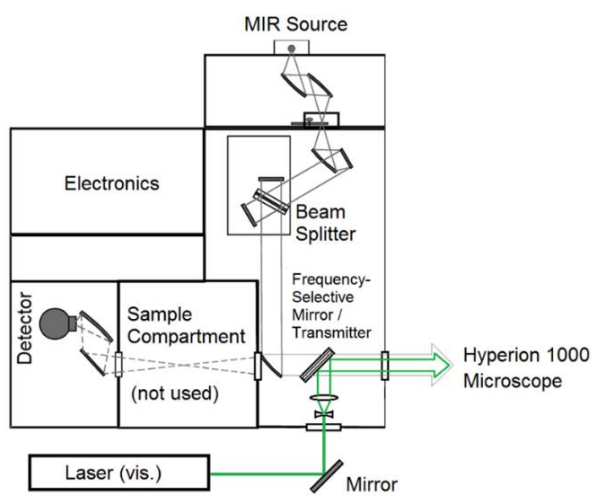

Fig. 2. Experimental setup for studying the infrared transmission of a plasmonic nanostructure composed of split ring resonators and its variation under illumination with visible light.

Hybrid aligned nematic (HAN) LC test cells are used, which consist of one substrate with homeotropic (= perpendicular) alignment and one substrate with planar parallel alignment of the LC director. Thin sandwich cells were fabricated by assembling two aligning substrates (from E.H.C. Ltd. Japan) separated by Mylar sheet spacers with a thickness of 50 µm. For all measurements, the HAN test cells were placed in the microscope with the homeotropically aligning substrate pointing towards the light source. The effect of addressing the MR-doped sample by the visible laser was studied through time-resolved measurements performed while switching the laser on and off. Major challenges of our opto-optic experiment were the proper positioning of the illuminated part of the metasurface and the large spectral deviation of the controlling visible laser beam and the probing infrared beam. In order to make sure that the entire part of the metasurface probed by the IR radiation can be illuminated with visible light, the focal point of the green laser beam was placed in front of the MR-doped LC layer thereby expanding at the position of the sample. The Cassegrain objective of the microscope turned out to impede this uniform illumination of the sample both because of geometrical reasons and because of the limited reflectivity of its gold mirrors in the visible range. Thus it was replaced by a transmitting microscope lens (20x, NA 0.35). Because of compromises that had to be made in order to facilitate the propagation of infrared radiation at lower wavenumbers, the intensities in the wavenumber range above 8000 cm$^{-1}$ are reduced, which causes a limited signal-to-noise ratio of spectra in this range. For time-resolved data recording, the FTIR spectrometer was operating in a fast scan mode. The control laser was started and stopped by a trigger signal from the spectrometer. Using a step-scan method was not necessary.

## 3. Results and discussion

The results of the baseline investigations (Fig. 3 – 5) show the NIR transmission of the HAN cell placed between crossed polarizers without and with additional illumination by the addressing visible laser. The transmission data given in figures 3 – 5 correspond to the intensity ratio $I_\perp/I_\parallel$, where $I_\perp$ and $I_\parallel$ correspond to the intensities transmitted through the polarizer, the LC sample and the analyzer, when the latter is crossed or parallel to the polarizer, respectively. Thus, zero transmission indicates that the LC sample has no influence on the state of polarization, while the maximum value of $I_\perp/I_\parallel = 1$ indicates a rotation of the plane of polarization by 90°. When the alignment direction of the planar alignment layer of HAN cell is aligned parallel (Fig. 3, Fig. 4) or perpendicular (Fig. 5) with respect to the plane of polarization of the incident radiation, no NIR radiation is transmitted in the ground state (laser off). This behavior is expected because the projection of the LC director on a plane perpendicular to the propagation direction of the electromagnetic waves points in the same direction at any position in the HAN cell; when this direction is oriented parallel or perpendicular to the plane of polarization, only either the extraordinary or the ordinary refractive index is apparent, respectively. Switching from the dark to a bright state appears due to the laser illumination [Fig. 3 (a), (b)]. If the sample is illuminated for less than 30 seconds, the dark state reappears, when the laser is switched off. A possible explanation for this behavior is illustrated in Fig. 3 (c). Due to a low energy dose of laser illumination, the SINE is expected to align the LC director perpendicular to the plane of polarization at the substrate that is close to the light source [Fig. 1 (c)]. This effect induces a twisted nematic (TN) structure, which is known for its waveguiding effect. Through this effect, the plane of polarization of the linearly polarized IR radiation is rotated by 90°, thereby enabling NIR transmission through the crossed analyzer. The switching behavior appears for significant different wavenumbers [Fig. 3 (a), (b)]. This observation confirms that the NIR intensity change is caused by a rotation of the plane of polarization (the waveguiding effect), not by an induced ellipticity of the NIR radiation.

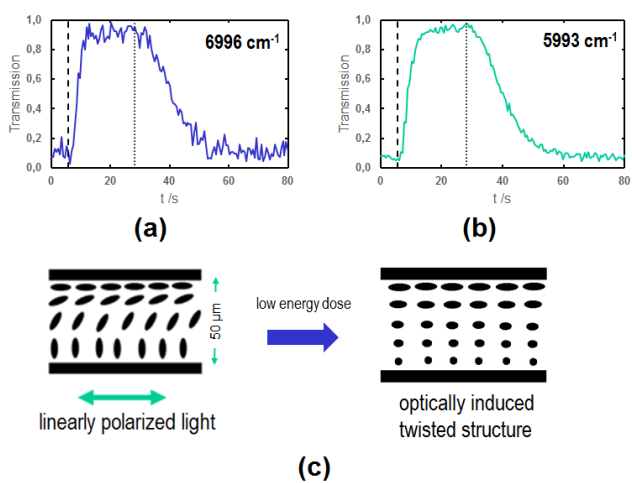

Fig. 3. Top: Transmission of a HAN cell of a liquid crystal doped with methyl red at (a) $\lambda^{-1}$ = 6996 cm$^{-1}$ and (b) $\lambda^{-1}$= 5993 cm$^{-1}$ under illumination with a low energy dose of visible, linearly polarized light (0.8 mW/mm²). Dashed line: time, when the laser was switched on; dotted line: time, when the laser was switched off. (c) Director field development, which explains the reversible effect that was observed.

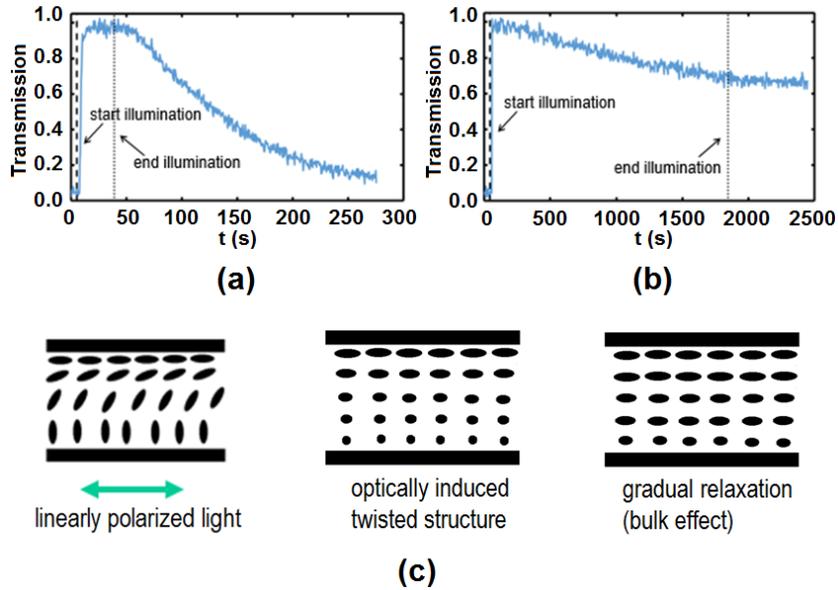

Fig. 4. Top: Long term development of the transmission of the hybrid aligned, dye-doped LC cell under illumination with visible, linearly polarized light (2.6 mW/mm²): (a) Illumination by 40 s results in a reversible change. (b) For the long term exposure, the transmission starts to decay gradually after 60 s and remains constantly non-zero after 30 minutes. (c) Director field development, which explains the observed behavior by a reversible surface effect during the first minute of illumination followed by an irreversible bulk effect.

After illumination with higher laser intensity for 40 seconds, relaxation into the dark state is still observed but requires more time [Fig. 4 (a)]. However, during a long illumination [Fig. 4 (b)] the intensity decreases gradually with time during illumination and remains constant after the laser illumination is switched off. This effect can be described by the SINE [Fig. 1 (c)] since a high energy dose is expected to align the LC director parallel to the linearly polarized laser illumination. This bulk effect gradually untwists the TN structure [Fig. 4 (c)], which might even end in a dark state after sufficiently long exposure time. Moreover, this bulk effect appearing at high energy doses of the illumination is reported to be irreversible, as observed in our experiment [Fig. 4 (b)]. To further illustrate this principle, the response of a HAN cell rotated by 90° was investigated [Fig. 5]. In contrast to the previous results [Fig. 4], no significant change of the transmission appears after a short illumination duration [Fig. 5 (a)]. There is only a small change in transmission during the illumination, which vanishes with ongoing illumination. A possible explanation for this behavior is illustrated in Fig. 5 (c). The condition of the low energy dose of illumination is fulfilled, which can be expected to cause a realignment from the HAN structure to an optically induced uniform structure. Most likely, the LC director rotates during the switching process owing to the elastic properties. In contrast, the intensity increases gradually with time during a long illumination and remains constant after the laser illumination is switched off. In this case, a TN structure is gradually formed [Fig. 5 (c)], while in the former case [Fig. 4], a TN structure was gradually untwisted during a long illumination due to the bulk effect. Furthermore, the observed long-term effect is, again in agreement with the literature, irreversible. After thorough baseline investigations of the underlying behavior of the MR-doped LC, a possible applicability for opto-optical use was investigated.

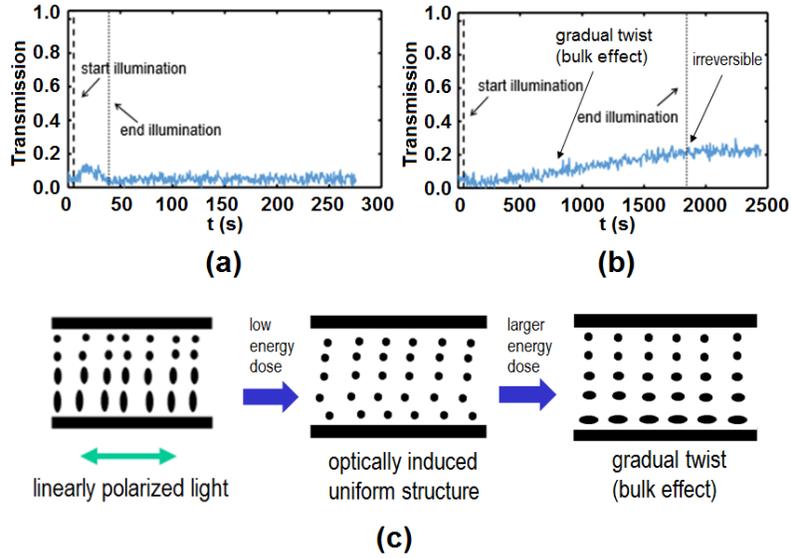

Fig. 5. Top: Development of the transmission under illumination with linearly polarized light (2.6 mW/mm²), when the azimuthal orientation of the hybrid aligned, dye-doped LC cell is rotated by 90° with respect to the plane of polarization: (a) Illumination for 40 s; (b) illumination for 30 minutes. (c) Director field development, which explains the observed behavior.

As a polarization-sensitive metasurface, SRRs were combined with an MR-doped LC cell [Fig. 6 (a)]. As previously described, SRRs possess polarization-dependent resonant modes due to their symmetry. The different IR transmission spectra of the sample [Fig. 6 (b)-(d)], which are measured before, during and after the illumination, illustrate a distinct opto-optical behavior of an MR-doped LC cell. Prior to illumination, the IR transmission spectrum of the SRR metasurface shows absorption at 4900 cm$^{-1}$, which can be attributed to the excitation of the first resonant mode, due to the orientation of the SRRs with respect to the plane of polarization of the incoming IR radiation. At the end of the laser illumination, absorption at 8600 cm$^{-1}$ is observed. Obviously, the second resonant mode is now excited, because the plane of polarization of the IR radiation is rotated by 90° owing to the optically induced TN structure. (The noise observed at wavenumbers of about 8000 cm$^{-1}$ and larger is caused by low radiation intensities in this spectral range, because the optical components in our experiment have to facilitate the propagation of both visible and infrared radiation, as explained in the experimental section). After the laser is switched off, the absorption at 4900 cm$^{-1}$ reappears. Obviously, the first resonant mode is excited again, because the plane of polarization of the IR radiation is rotated back, which indicates reversible relaxation of the optically induced TN structure back to the initial configuration.

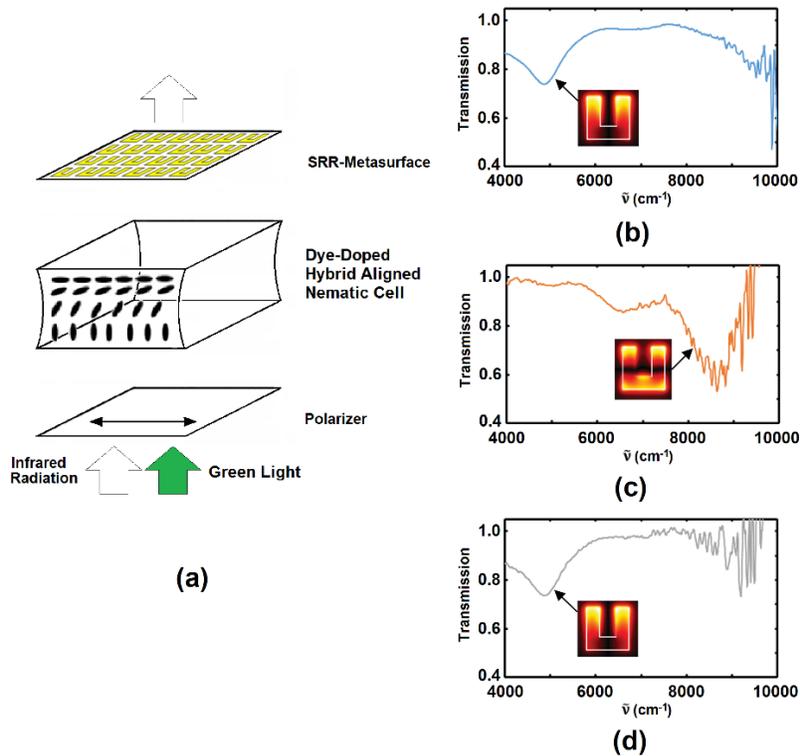

Fig. 6. Investigation of the split ring resonator (SRR) array illuminated through the light-controlled dye-doped LC cell. (a) Experimental geometry. (b-d) IR transmission spectra of the sample (b) before, (c) during, and (d) after illumination (2.6 mW/mm², 22.5 s).

## 4. Conclusion

Summarizing, we have studied the baseline opto-optical response of an MR-doped LC cell with the initial HAN configuration and also the possibility of an application for opto-optical devices, exemplified with an SRR metasurface. The results of the baseline investigations can be explained very well by the specific features of the SINE described in the literature [50]. Depending on the energy dose of the laser radiation, the LC molecules located at the substrate closer to the source of illumination are orientated in different manners. Different optically induced structures are achieved owing to the elastic properties of the LC and its alignment at the counter substrate. An optically induced TN-structure is achieved in the following cases. Case 1: short laser illumination and an LC alignment at the counter substrate parallel to the plane of polarization of the incoming radiation. Case 2: long laser illumination and an LC alignment at the counter substrate perpendicular to the plane of polarization of the incoming radiation (gradually formed). However, an optically induced uniform structure emerges, when the LC alignment at the counter substrate is perpendicular to the cases 1 and 2 described above. The relaxation times depend essentially on the energy doses. In the low energy dose regime, the optical relaxation times increase with increasing energy dose, in agreement with earlier studies [55]. At a very high energy dose, the optical switching effect becomes irreversible, which is also consistent with the literature [55]. Our work demonstrates that the described effects can be applied for an opto-optical device that is based on a polarization-sensitive metasurface composed of SRRs. Laser illumination yields a reversible switching between the excited resonant modes.

In contrast to continuous shifts of the resonance wavenumbers, which can be achieved by using the liquid crystal only as a dielectric environment of the resonators [27, 37], the control of the state of polarization of the IR radiation applied here enables selective addressing of different resonant modes, the wavenumbers of which differ by more than 3500 cm$^{-1}$. The high transmission maximum achieved for reversible switching of the pure liquid crystal (Fig. 3) indicates that the liquid crystal cell rotates the plane of polarization efficiently by 90°, as expected. A comparison with earlier electro-optic studies on SRR arrays [27, 29] reveals that the optically induced contrast ratio $I_{max}/I_{min}$ achieved for the transmission of the metasurface (Fig. 6: $I_{max}/I_{min} \approx 1.7$ at 4920 cm$^{-1}$ and $I_{max}/I_{min} \approx 2.8$ at 8540 cm$^{-1}$, respectively) is limited by the absorption of the SRR array. Since the optical polarization control can be readily applied to other polarization-sensitive metasurfaces, we expect that much larger contrasts can be achieved. The switching times of the reversible effect observed here (Fig. 3: $\tau_{on} \approx$ 4-5 s and $\tau_{off} \approx$ 20 s) may be further reduced down to $\approx$ 100 ms under cw illumination [56] and to a few ms for excitation with ns-pulses [52]. So, fast all-optical modulation [18, 22] is not expected for the SINE effect, but bi-stable switching might be envisioned. The small speed of the effect is compensated by the very low intensity 0.8 mW/mm$^2$ needed for cw excitation in comparison with ns-pulses (100-250 W/mm$^2$) [22] and fs-pulses ($3 \cdot 10^7$ W/mm$^2$) [18]. Moreover, the variety of nonlinear optical effects in LCs [48-51] and photo-sensitive materials [57-59] promise further opportunities of optical addressing at visible and ultraviolet frequencies, including the application of monomer and polymer liquid crystals with azobenzene moieties [57], both for surface effects ('command surfaces') [58] and bulk effects [59]. For example, using a polymer azobenzene, a nematic-isotropic phase transition was induced in 200 ns by applying a ps-pulse at $\lambda$ = 355 nm [59]. Thus, the opportunity of optical addressing is expected to have a large impact on a variety of opto-optical effects in metasurfaces that are related to polarization. To mention just one example, holograms and holographic optical elements utilizing Pancharatnam-Berry metasurfaces can be expected to become addressable not only by electrical [60] but also by optical signals. Based on the progress described in Refs. [37] and [38], we are confident, that our concept of optical addressing can also be applied to less lossy dielectric metasurfaces.

## 5. Funding, acknowledgments, and disclosures

### 5.1. Funding

Financial support by the German Research Foundation (DFG: GRK 1464) is gratefully acknowledged. This project has received funding from the European Research Council (ERC) under the European Union's Horizon 2020 research and innovation programme (grant agreement No 724306).
.

### 5.2. Acknowledgments

The authors would like to thank Liana Lucchetti for illuminating discussions about the surface-induced nonlinear effect and Cedrik Meier for providing access to his electron beam lithography system.

## 6. References

1. A. Sihvola, S. Tretyakov and D. de Baas, "Metamaterials with Extreme Material Parameters," J. Commun. Technol. El. **52**(9), 986-990 (2007).
2. W. Cai and V.M. Shalaev, *Optical Metamaterials* (Springer-Verlag, 2010).
3. D.R. Smith, S. Schultz, P. Markos and C.M. Soukoulis, "Determination of effective permittivity and permeability of metamaterials from reflection and transmission coefficients," Phys. Rev. B **65**, 195104 (2002).
4. V.G. Veselago, "The electrodynamics of substances with simultaneously negative values of ε and μ," Sov. Phys. Usp. **10**, 509-514 (1968).

5. R.A. Shelby, D.R. Smith and S. Schultz, "Experimental Verification of a Negative Index of Refraction," Science **292**, 77–79 (2001).
6. J.B. Pendry, "Negative Refraction Makes a Perfect Lens," Phys. Rev. Lett. **85**, 3966–3969 (2000).
7. Z.W. Liu, H. Lee, Y. Xiong, C. Sun and X. Zhang, "Far-field optical hyperlens magnifying sub-diffraction-limited objects," Science **315**, 1686 (2007).
8. D. Schuring, J. Mock, B. Justice, B. Cummer, J. Pendry, A. Starr and D. Smith, "Metamaterial Electromagnetic Cloak at Microwave Frequencies," Science **314**, 977-980 (2006).
9. M. Kadic, G. W. Milton, M. van Hecke and M. Wegener, "3D metamaterials", Nature Rev. Phys. **1**, 198–210 (2019).
10. H.-T. Chen, A. J, Taylor and N. Yu, "A review of metasurfaces: physics and applications", Rep. Prog. Phys.**79**, 076401 (2016).
11. S. M. Kamali, E. Arbabi, A. Arbabi and A. Faraon, "A review of dielectric optical metasurfaces for wavefront control", Nanophotonics **7**, 1041–1068 (2018).
12. Y. Kivshar, "All-dielectric meta-optics and non-linear nanophotonics", Nat. Sci. Rev. **5**, 144–158 (2018).
13. O. Hemmatyar, S. Abdollahramezani, Y. Kiarashinejad, M. Zandehshahvar and A. Adibi, "Full color generation with Fano-type resonant $HfO_2$ nanopillars designed by a deep-learning approach", Nanoscale **11**, 21266-21274 (2019).
14. M. Khorasaninejad, W. T. Chen, R. C. Devlin, J. Oh, A. Y. Zhu and F. Capasso, "Metalenses at visible wavelengths: Diffraction-limited focusing and subwavelength resolution imaging", Science **352**, 1190-1194 (2016).
15. S. Colburn, A. Zhan and A. Majumdar, "Varifocal zoom imaging with large area focal length adjustable metalenses," Optica **5**, 825-831 (2018).
16. A. I. Holsteen, S. Raza, P. Fan, P. G. Kik and M. L. Brongersma, "Purcell effect for active tuning of light scattering from semiconductor optical antennas", Science **358**, 1407-1410 (2017).
17. T. Lewi, P. P. Iyer, N. A. Butakov, A. A. Mikhailovsky and J. A. Schuller, "Widely Tunable Infrared Antennas Using Free Carrier Refraction", Nano Lett. **15**, 8188-8193 (2015).
18. M. R. Shcherbakov, S. liu, V. V. Zubyuk, A. Vaskin, P. P. Vabishchevich, G. Keeler, T. Pertsch, T. V. Dolgova, I. Staude, I. Brener and A. A. Fedyanin: "Ultrafast all-optical tuning of direct-gap semiconductor metasurfaces", Nature Commun. **8**, 17 (2017).
19. T. Lewi, H. A. Evans, N. A., Butakov and J. A. Schuller, "Ultrawide Thermo-optic Tuning of PbTe Meta-Atoms", Nano Lett. **17**, 3940-3945 (2017).
20. A. Howes, W. Wang, I. Kravchenko and J. Valentine, "Dynamic transmission control based on all-dielectric Huygens metasurfaces", Optica **5**, 787-792 (2018).
21. G. K. Shirmanesh, R. Sokhoyan, R. A. Pala and H. A. Atwater, "Dual-Gated Active Metasurface at 1550 nm with Wide (>300°) Phase Tunability", Nano Lett. **18**, 2957-2963 (2018).
22. B. Gholipour, J. Zhang, K. F. MacDonald, D. W. Hewak and N. I. Zheludev, "An all-optical, non-volatile, bidirectional, phase-change meta-switch", Adv. Mater. **25**, 3050-3054 (2013).
23. Z. L. Sámson, K. F. MacDonald, F. De Angelis, B. Gholipour, K. Knight, C. C. Huang, E. Di Fabrizio, D. W. Hewak, and N. I. Zheludev. "Metamaterial electro-optic switch of nanoscale thickness", Appl. Phys. Lett. **96**, 143105 (2010).
24. N. A. Butakov, I. Valmianski, T. Lewi, C. Urban, Z. Ren, A. M. Mikhailovsky, S. D. Wilson, I. K. Schuller and J. A. Schuller, "Switchable Plasmonic–Dielectric Resonators with Metal–Insulator Transitions", ACS Photonics **5**, 371-377 (2018).
25. N. A. Butakov, M. W. Knight, T. Lewi, P. P. Iyer, D. Higgs, H. T. Chorsi, J. Trastoy, J. Del Valle Granda, I. Valmianski, C. Urban, Y. Kalcheim, P. Y. Wang, P. W. C. Hon, I. K. Schuller and J. A. Schuller, "Broadband electrically tunable dielectric resonators using metal-insulator transitions", ACS Photonics **5**, 4056-4060 (2018).
26. Z. Zhu, P. G. Evans, R. F. Haglund Jr. and J. G. Valentine, "Dynamically reconfigurable metadevice employing nanostructured phase-change materials", Nano Lett. 17, 4881-4885 (2017).
27. B. Atorf, H. Mühlenbernd, M. Muldarisnur, T. Zentgraf and H. Kitzerow, "Effect of alignment on a liquid crystal / split ring resonator metasurface," Chem. Phys. Chem. **15**, 1470-1476 (2014).
28. F. Zhang, L. Kang, Q. Zhao, J. Zhou, X. Zhao and D. Lippens, "Magnetically tunable left handed metamaterials by liquid crystal orientation," Opt. Express **17**(6), 4360-4366 (2009).
29. B. Atorf, H. Mühlenbernd, M. Muldarisnur, Z. Zentgraf and H. Kitzerow, "Electro-optic tuning of split ring resonators embedded in a liquid crystal," Opt. Lett. **39**(5), 1129-1132 (2014).
30. A.C. Tasolamprou, D.C. Zografopoulos and E.E. Kriezis, "Liquid crystal-based dielectric loaded surface plasmon polariton optical switches," J. Appl. Phys. **110**, 093102 (2011).
31. D.H. Werner, D.H. Kwon, I.C. Khoo, A.V. Kildishev and V.M. Shalaev, "Liquid crystal clad near-infrared metamaterials with tunable negative-zero-positive refractive indices," Opt. Express **15**(6), 3342-3347 (2007).
32. Y.J. Liu, Q. Hao, J.S.T. Smalley, J. Liou, I.C. Khoo and T.J. Huang, "A frequency-addressed plasmonic switch based on dual-frequency liquid crystals," Appl. Phys. Lett. **97**, 091101 (2010).
33. Q. Hao, Y. Zhao, B.K. Juluri, B. Kiraly, J. Liou, I.C. Khoo and T.J. Huang, "Frequency addressed tunable transmission in optically thin metallic nanohole arrays with dual frequency liquid crystals," J. Appl. Phys. **109**, 084340 (2011).
34. Q. Zhao, L. Kang, B. Du, B. Li, J. Zhou, H. Tang, X. Liang and B. Zhang, "Electrically tunable negative permeability metamaterials based on nematic liquid crystals," Appl. Phys. Lett. **90**, 011112 (2007).

35. F. Zhang, Q. Zhao, W. Zhang, J. Sun, J. Zhou and D. Lippens, "Voltage tunable short wire-pair type of metamaterial infiltrated by nematic liquid crystal," Appl. Phys. Lett. **97**, 134103 (2010).
36. O. Buchnev, J. Wallauer, M. Walther, M. Kaczmarek, N.I. Zheludev and V.A. Fedotov, "Controlling intensity and phase of terahertz radiation with an optically thin liquid crystal-loaded metamaterial," Appl. Phys. Lett. **103**, 141904 (2013).
37. A. Komar, Z. Fang, Justus Bohn, J. Sautter, M. Decker, A. Miroshnichenko, T. Pertsch, I. Brener, Y. S. Kivshar, I. Staude and D. N. Neshev, "Electrically tunable all-dielectric optical metasurfaces based on liquid crystals", Appl. Phys. Lett. **110**, 071109 (2017).
38. S.-Q. Li, X. Xu, R. M. Veetil, V. Valuckas, R. Paniagua-Domínguez and A. I. Kuznetsov, "Phase-only transmissive spatial light modulator based on tunable dielectric surface", Science **364**, 1087-1090 (2019).
39. P.G. De Gennes and J. Prost, *The Physics of Liquid Crystals* (Oxford University, 1993).
40. M. Schadt and W. Helfrich, "Voltage-dependent Optical Activity of a Twisted Nematic Liquid Crystal," Appl. Phys. Lett. **18**, 127−128 (1971).
41. S.-T. Wu, U. Efron, L.D. Hess, "Optical rotatory power of 90° twisted nematic liquid crystals," Appl. Phys. Lett. **44**, 842−844 (1984).
42. M. Schadt, "Milestone in the history of field-effect liquid crystal displays and materials," Jpn. J. Appl. Phys. **48**, 03B001 (2009).
43. W. Kaminsky, "Experimental and phenomenological aspects ofcircular birefringence and related properties in transparent crystals," Rep. Prog. Phys. **63**, 1575−1640 (2000).
44. D.R. Smith, W.J. Padilla, D.C. Vier, S.C. Nemat-Nasser and S. Schultz, "Composite Medium with Simultaneously Negative Permeability and Permittivity," Phys. Rev. Lett. **84**(18), 4184-4187 (2000).
45. C. Rockstuhl, T. Zentgraf, H. Guo, N. Liu, C. Etrich, I. Loa, K. Syassen, J. Kuhl, F. Lederer and H. Giessen, "Resonances of split-ring resonator metamaterials in the near infrared," Appl. Phys. B **84**, 219-227 (2006).
46. C. Rockstuhl, F. Lederer, C. Etrich, T. Zentgraf, J. Kuhl and H. Giessen, "On the reinterpretation of resonances in split-ring-resonators at normal incidence," Opt. Express **14**(19), 8827-8836 (2006).
47. P. B. Johnson and R. W. Christy, "Optical Constants of the Noble Metals," Phys. Rev. B **6**, 4370-4379 (1972).
48. B.Ya. Zel'dovich, N.F. Pilipetski, A.V. Sukhov and N.V. Tabiryan, "Giant optical nonlinearity in the mesophase of a nematic liquid crystal (NCL)," JETP Lett. **31**, 263-267 (1980).
49. I. Khoo, "Nonlinear optics of liquid crystalline materials," Phys. Rep. **471**, 221-267 (2009).
50. I. Jánossy, A.D. Lloyd and B.S. Wherrett, "Anomalous Optical Freedericksz Transition in an Absorbing Liquid Crystal," Mol. Cryst. Liq. Cryst. **179**, 1-12 (1990).
51. I.C. Khoo, S. Slussarenko, B.D. Guenther, M.Y. Shih, P.H. Chen and W.V. Wood, "Optically induced space-charge fields, dc voltage, and extraordinarily large nonlinearity in dye-doped nematic liquid crystals," Opt. Lett. **23**, 253-255 (1998).
52. F. Simoni, L. Lucchetti, D. E. Lucchetta and O. Francescangeli, "On the origin of the huge nonlinear response of dye-doped liquid crystals," Opt. Express **9**(2), 85-90 (2001).
53. A. Petrossian and S. Residori, "Surfactant enhanced reorientation in dye-doped nematic liquid crystals," Europhys. Lett. **60**(1), 79–85 (2002).
54. L. Lucchetti, M. Di Fabrizio, O. Francescangeli and F. Simoni, "Colossal optical nonlinearity in dye doped liquid crystals," Opt. Commun. **233**, 417-424 (2004).
55. L. Lucchetti and F. Simoni, "Role of space charges on light-induced effects in nematic liquid crystals doped by methyl red," Phys. Rev. E **89**, 032507 (2014).
56. L. Lucchetti, M. Gentili and F. Simoni, "Effects leading to colossal optical nonlinearity in dye-doped liquid crystals", IEEE Journal of Selective Topics in Quantum Electronics **12**(3), 422-430 (2006).
57. Y, Yu and T. Ikeda, "Alignment modulation of azobenzene-containg liquid crystal systems by photochemical reactions", J. Photochem. Photobiol. C: Photochemistry Reviews **5**, 247-265 (2004).
58. K. Ichimura, "Photoalignment of Liquid-Crystal Systems", Chem. Rev. **100**, 1847-1873 (2000).
59. T. Ikeda, "Photochemical Modulation of Refractive Index by Means of Photosensitive Liquid Crystals", Mol. Cryst. Liq. Cryst. 364, 187-197 (2001).
60. B. Atorf, H. Rasouli, H. Mühlenbernd, B. Reineke, T. Zentgraf and H. Kitzerow, "Switchable Plasmonic Hologram Utilizing the Electrooptic Effect of a Liquid Crystal Circular Polarizer," J. Phys. Chem. C **122**(8), 4600-4606 (2018).